\documentclass[aps,preprint,superscriptaddress,nofootinbib]{revtex4-1}
\pdfoutput=1

\usepackage[T1]{fontenc}
\usepackage[utf8]{inputenc}

\usepackage{amsmath, amssymb, amsfonts, bm}
\usepackage{physics}

\usepackage{graphicx}
\usepackage{float}
\usepackage{wrapfig}
\usepackage{booktabs}
\usepackage{placeins}
\usepackage{tabularx}
\usepackage{caption}
\usepackage{subcaption}  

\usepackage{algorithm}
\usepackage{algpseudocode}

\usepackage{xcolor}
\usepackage{comment}
\usepackage{lipsum}  

\usepackage[colorlinks=true,
            linkcolor=blue,
            citecolor=blue,
            filecolor=blue,
            urlcolor=blue]{hyperref}

\DeclareUnicodeCharacter{2212}{-}

\newcommand{\eq}[1]{Eq.~(\ref{#1})}

\newcommand{\fig}[1]{Fig.~\ref{#1}}

\newcommand{\HHE}{\hat{H}_{\rm eff}}
\newcommand{\HH}{\hat{H}}
\def\bea{\begin{eqnarray}}
\def\eea{\end{eqnarray}}

\begin{document}

\title{Estimating Trotter Approximation Errors to Optimize Hamiltonian Partitioning for Lower Eigenvalue Errors} 

\author{Shashank G. Mehendale$^*$}
\affiliation{Chemical Physics Theory Group, Department of Chemistry, University of Toronto, Toronto, Ontario M5S 3H6, Canada}
\affiliation{Department of Physical and Environmental Sciences, University of Toronto Scarborough, Toronto, Ontario M1C 1A4, Canada}

\author{Luis A. Martínez-Martínez$^*$}
\affiliation{Chemical Physics Theory Group, Department of Chemistry, University of Toronto, Toronto, Ontario M5S 3H6, Canada}
\affiliation{Department of Physical and Environmental Sciences, University of Toronto Scarborough, Toronto, Ontario M1C 1A4, Canada}

\author{Prathami Divakar Kamath$^*$}
\affiliation{Department of Metallurgical Engineering and Materials Science, Indian Institute of Technology Bombay, Maharashtra, 400076, India}
\author{Artur F. Izmaylov}
\affiliation{Chemical Physics Theory Group, Department of Chemistry, University of Toronto, Toronto, Ontario M5S 3H6, Canada}
\affiliation{Department of Physical and Environmental Sciences, University of Toronto Scarborough, Toronto, Ontario M1C 1A4, Canada}

\def\thefootnote{*}\footnotetext{These authors contributed equally to this work}\def\thefootnote{\arabic{footnote}}

\begin{abstract}

Trotter approximation in conjunction with Quantum Phase Estimation can be used to extract eigen-energies of a many-body Hamiltonian on a quantum computer. There were several ways proposed to assess the quality of this approximation based on estimating the norm of the difference between the exact and approximate
evolution operators. Here, we explore how different error estimators correlate with the true error in the ground state energy due to Trotter approximation. For a set of small molecules we calculate these exact error in ground-state electronic energies due to the second-order Trotter approximation. Comparison of these errors with previously used upper bounds show correlation less than 0.4 across various Hamiltonian 
partitionings. On the other hand, building the Trotter approximation error estimation based on perturbation theory up to a second order in the time-step for eigenvalues provides estimates with very good correlations with the exact Trotter approximation errors. These findings highlight the non-faithful character of norm-based estimations for prediction of a Trotter-based eigenvalue estimation performance and the need of alternative estimators. The developed perturbative estimates can be used for practical time-step 
and Hamiltonian partitioning selection protocols, which are needed for an accurate assessment of quantum resources.
\end{abstract}

\maketitle

\section{Introduction}

Solving the electronic structure problem is one of the anticipated uses of quantum computing.
As an eigenvalue problem with a Hamiltonian operator that can be expressed compactly, 
this problem is convenient for quantum computing because classical-quantum 
data transfer is usually a bottleneck.\cite{hoefler2023disentangling}
Obtaining electronic wavefunctions and energies is one of the key procedures in 
first principles modeling of molecular physics 
since molecular energy scale is dominated by the electronic part. 
Yet, solving this problem scales exponentially with the size unless some approximations are made.

Fault-tolerant quantum computers offer potential advantages
for efficient estimation of energy eigenvalues through exponential
speedup with respect to classical methods, by means of the Quantum
Phase Estimation (QPE) algorithm \cite{osti_1827672}.
The QPE framework contains three main parts: 1) initial state preparation, 
2) procedure for an evolution or a walker operator that involves the Hamiltonian encoding, 
and 3) the eigenvalue extraction. Here, we focus on the second part, 
two main approaches for the Hamiltonian encoding are representing the Hamiltonian 
exponential function via the Trotter approximation and embedding the Hamiltonian as 
a block of a larger unitary via decomposing the Hamiltonian as a 
Linear Combination of Unitaries (LCU).

Within the Trotter approximation, the target Hamiltonian is decomposed
into easy-to-simulate (or fast-forwardable) Hamiltonian fragments:
\begin{equation}
\hat{H}=\sum_{m=1}^{M}\hat{H}_{m}\label{Gen_Decomp}
\end{equation}
\color{black} and the exact unitary evolution operator for an arbitrary simulation time $\tau$ is approximated using the time evolution of the fragments $\hat{H}_m$. The second-order Trotter approximation is given by

\begin{equation}
\hat{U}(\tau)=e^{-i\tau\hat{H}}\approx
\left(\prod_{m=1}^{M}e^{-i\hat{H}_{m}\tau/(2n)} \prod_{m=M}^{1}e^{-i\hat{H}_{m}\tau/(2n)} \right)^n =
\left(\hat{U}^{(2)}_{T}(\tau/n)\right)^{n}\label{trotter_2nd}
\end{equation}
where the approximation is exact up to second order in $\tau$.
\color{black}
This approximate representation of the exact time evolution operator
introduces a deviation in the spectrum of the simulated time evolution
unitary with respect to the exact one. For estimation of energy eigenvalues
through QPE under a fixed target error, it is therefore crucial to
rationalize the scaling of this deviation with the time scale used
for discretization of the total simulation time as well as its dependence
with different Hamiltonian partitioning schemes.

In spite of less favorable time scaling of the Trotter approach compared to the LCU based techniques, it has benefits of a lower ancilla qubit overhead and possibility for using commutation relation between terms of the Hamiltonian for formulating fast-forwardable fragments.  
Yet, one difficulty for practical use of the Trotter approximation is estimation of its error. 
This estimation is needed for choosing the evolution time-step and the overall error estimation. 
Another use of the Trotter approximation error is choosing the Hamiltonian partitioning that 
minimizes error and thus the number of steps required.    


Recently, upper bounds were formulated for the norm of the difference between propagators, 
\begin{gather}
||\hat U(\tau) - \hat{U}^{(2)}_{T}(\tau/n)^{n}|| \le \frac{\alpha \tau^3}{n^2}, \label{SpecNDiff}\\
\alpha = \frac{1}{12} \sum_{m_1 = 1}^{M} \left\| \Big[ \sum_{m_3 = m_1 + 1}^{M} H_{m_3}, \Big[ \sum_{m_2 = m_1 + 1}^{M} H_{m_2}, H_{m_1} \Big] \Big] \right\| \nonumber \\
+ \frac{1}{24} \sum_{m_1 = 1}^{M} \left\| \Big[ H_{m_1}, \Big[ H_{m_1}, \sum_{m_2 = m_1 + 1}^{M} H_{m_2} \Big] \Big] \right\|. \label{eq:alpha}
\end{gather}
which allowed one to estimate the effect of the Trotter approximation on the accuracy of 
dynamics \cite{childs2021Trot}. These estimates can be used to derive upper bounds for the energy 
error in QPE \cite{Reiher_2017}. In what follows, for brevity, we will refer to the time step as $t=\tau/n$.
However, it is known in general that the Trotter upper bounds are relatively loose and 
using them could lead to underestimation of appropriate time-step \cite{poulin2014trotter}. 
Considering that with some simplifications $\alpha$ values can be evaluated and used 
to differentiate various Hamiltonian partitionings \cite{MartinezMartinez2023assessmentofvarious},
it is interesting to examine how accurate $\alpha$-based trends are
compared to those using the exact Trotter approximation error in eigenvalues.

Here, we investigate using the exact error calculation for small systems whether 
the Trotter approximation error upper bounds can be used to differentiate Hamiltonian partitionings. 
We also explore alternative estimates of the Trotter approximation error for eigenvalues 
based on time-independent perturbation theory. Such theory can be 
built by representing the Trotter propagator as 
\bea\label{eq:Heff}
\hat{U}^{(2)}_{T}(t)=e^{-it\hat{H}_{\rm eff}(t)}
\eea
and performing 
perturbative analysis of the $\hat{H}_{\rm eff}(t)$ spectrum. 
Even though perturbative estimates are not upper bounds, 
they can be used for differentiating between various Hamiltonian partitioning schemes. 
As for predicting the Trotter step, one can use perturbative estimates as a 
first step in the iterative procedure suggested recently.\cite{rendon2022}

\section{Perturbative error estimates \textbf{\label{PT_rat}}}
Time-independent perturbation theory is built by considering Baker-Campbell–Hausdorff
expansion of the second order Trotter evolution operator in \eq{eq:Heff}
\bea
\HHE(t) = \HH + \sum_k \hat{V}_k t^k. 
\label{eq: Heff_pert_series}
\eea
\textcolor{black}{By construction [see Eq. \eqref{trotter_2nd}], $\hat{U}_T^{(2)}(t)\hat{U}_T^{(2)}(-t) = 1$, implying $\hat{H}_{\text{eff}}(t) = \hat{H}_{\text{eff}}(-t)$. Therefore, only even order $\hat{V}_k$'s survive in Eq. \eqref{eq: Heff_pert_series}. The leading term is then given by [see Appendix \ref{Eff_BCH}]}
\begin{equation}\label{Pert_defs}
    \begin{split}
    \hat{V}_{2}&=-\frac{1}{24}\sum_{v'=v}^{2M}\sum_{v=\mu+1}^{2M}\sum_{\mu=1}^{2M-1}\left(1-\frac{\delta_{v',v}}{2}\right)[\hat{H}_{v'},[\hat{H}_{v},\hat{H}_{\mu}]],
   \end{split}
\end{equation}
where $H_{M+i}$ = $H_{M+1-i}$ for $i=1$ to $M$.
Note that in spite of $t$ dependence of $\HHE$, we do not need time-dependent perturbation theory 
since we are interested in eigenvalues of $\HHE$ as a function of $t$. 
Eigenvalues of $\HHE$ can be obtained as perturbative series starting from 
those of $\hat{H}$. Focusing on the ground state energy $E_0$, the correction from first-order perturbation theory can be written as
\[
E_{\text{GS}}^{(1)}=\langle\phi_{0}|\hat{V}_{2}|\phi_{0}\rangle t^{2},
\]
where $|\phi_{0}\rangle$ is the electronic ground state. Note that next correction to energy will be fourth order in time. This implies, the ground state energy of $H_{\text{eff}}$ is $E_{0}^{(T)} = E_{0}+\varepsilon_2 t^{2} + \mathcal{O}(t^4)$, where
\begin{equation}
\varepsilon_2=\langle\phi_{0}|\hat{V}_{2}|\phi_{0}\rangle. \label{eq:epsGS}
\end{equation}
Calculating $\varepsilon_2$ requires knowledge of the ground state of $\hat{H}$. Since it is not accessible for a general Hamiltonian, we approximate $\varepsilon_{2}$ using approximate eigenstate $|\psi_{0}\rangle$ obtained via cost efficient classical methods. We can then define an approximation to $\varepsilon_2$
given by 
\begin{equation}
\varepsilon_{\text{app}}=\langle\psi_{0}|\hat{V}_{2}|\psi_{0}\rangle.\label{eps_PT}
iv\end{equation}
The difference $|\varepsilon_2-\varepsilon_{\text{app}}|$ is expected to become smaller with larger overlap $|\langle\phi_{0}|\psi_{0}\rangle|$.\\

\color{black}

\section{Results and discussion\textbf{\label{Results}}}

Here, we assess correlations between the exact Trotter approximation errors 
and estimates based on $\alpha$ [\eq{eq:alpha}] and 
perturbative expression $\varepsilon_{\text{app}}$ [Eqs. (\ref{eps_PT})]. 
The Trotter approximation errors are obtained for electronic Hamiltonians of small molecules 
(H$_{2}$, LiH, BeH$_{2}$, H$_{2}$O, and NH$_{3}$) and various Hamiltonian partitioning schemes described in Appendix \ref{Methods}. The approximate ground state $\ket{\psi_0}$ is obtained from CISD calculations. For H$_{2}$, LiH, and BeH$_{2}$, the bond length is chosen to be 1 $\r{A}$. For H$_2$O and NH$_3$, the bond length is chosen to be 1.9 $\r{A}$. The overlap $|\braket{\phi_0}{\psi_0}|^2$ equals $93\% $ for H$_2$O and $83\%$ for NH$_3$. The exact Trotter approximation errors $|\Delta E_{T}|=|E^{(T)}_{0}-E_{0}|$ are computed by
numerical diagonalization of $\hat{H}$ and $\hat{H}_{\text{eff}}$ [\eq{eq:Heff}] as described in Appendix \ref{AutoCorr}.

\subsection{Exact Trotter approximation errors}
\begin{figure*}
\centering

\includegraphics[width=0.9\textwidth]{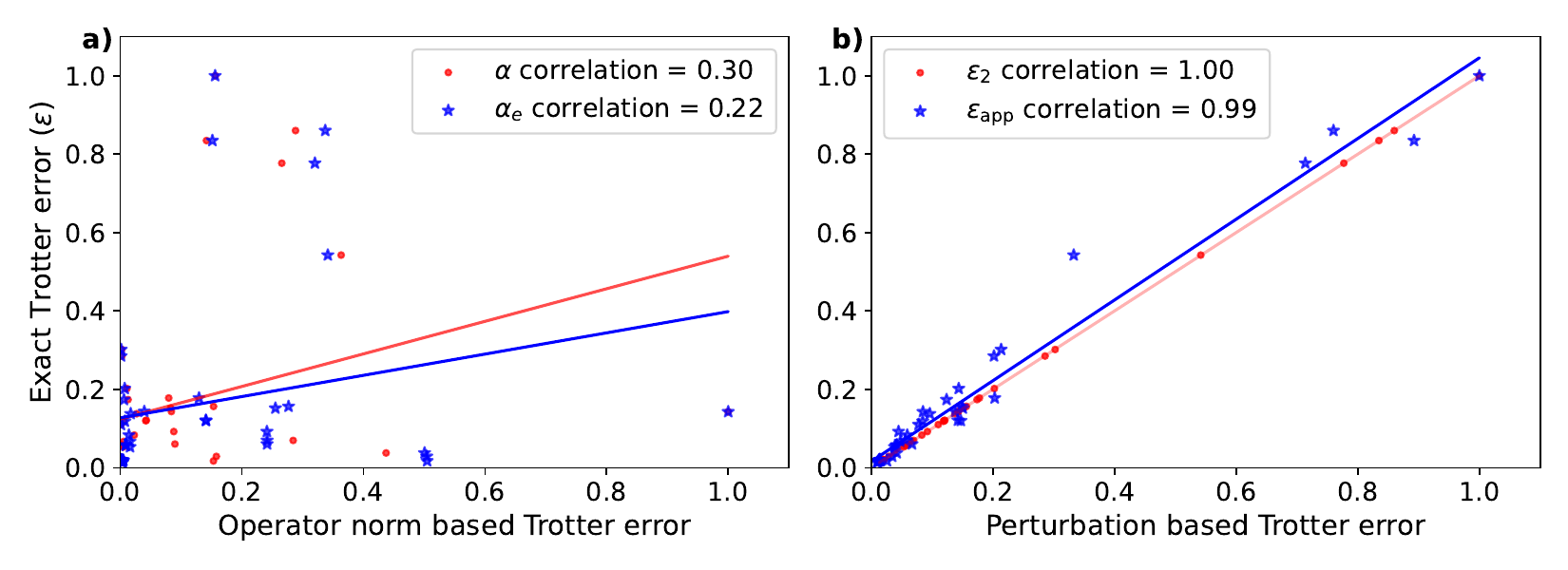}

\caption{ Correlation between exact Trotter error and the error estimators. Each point on the plot corresponds to a unique pair of molecule and fragmentation technique. The straight lines are plotted using the Pearson correlation coefficient, given in the legend. All axes are normalized to one. (a) $\varepsilon$ vs $\alpha$ correlation is denoted in red, while $\varepsilon$ vs $\alpha_e$ is denoted in blue. (b) $\varepsilon$ vs $\varepsilon_2$ correlation is denoted in red, while $\varepsilon$ vs $\varepsilon_{\text{app}}$ in blue. The corresponding numerical data is presented in appendix \ref{EpsTables}.}
\label{fig_sim}
\end{figure*}

We define $\varepsilon = \Delta E_T / t^2$ to represent the exact Trotter approximation error and examine the correlations between $\varepsilon$ and the error estimators. Comparing true errors with predictions based on $\alpha$ upper bound in \fig{fig_sim} (a) (red marker) shows poor correlation. Thus, it is not possible to determine the Hamiltonian partitioning 
performance in terms of the Trotter approximation error based on $\alpha$ values.  
Upper bounds based on $\alpha$'s are usually very loose, so we have considered an $\alpha$-like estimator
\begin{equation}\label{alphExc}
    \alpha_e=||\hat{U}(t) - \hat{U}_{T}^{(2)}(t)||/t^3.
\end{equation}
$\alpha_e$ captures the exact error in the time propagator introduced by the Trotter approximation. However, the correlation plot of \fig{fig_sim} (a) (blue marker) shows that $\alpha_e$ is also poorly correlated with $\varepsilon$. This discrepancy can be understood as a consequence of $\alpha$ and $\alpha_e$ being worst-case scenario metrics for the \textcolor{black}{deviation} (with respect to exact unitary propagation) that ensue from the Trotter approximation rather than a measure of deviation
with respect to the eigenspectrum of the target simulated Hamiltonian.\\

On the other hand, since $|\Delta E_{T}|=\varepsilon_2 t^2 +\mathcal{O}(t^4)$, 
for small enough $t$, $\varepsilon_2$ should exactly capture $\varepsilon$. We can observe this in \fig{fig_sim} (b), where we plot the correlation between $\varepsilon$ and $\varepsilon_2$ in red. The correlation coefficient, as expected, is one. In the same subplot, we also show the correlation between $\varepsilon$ and $\varepsilon_{\text{app}}$. We can see a strong correlation between the two quantities, with a 
correlation coefficient of 0.99. This suggests that despite having approximate wavefunctions with a fidelity as low as $0.83$, one can still predict the best and worst partitioning techniques.\\

\color{black}

\FloatBarrier

\subsection{Resource efficiency}\label{Res_Eff}

\begin{table*}
\resizebox{\textwidth}{!}{
\begin{tabular}{|c|c|c|c|c|c|c|}
\hline 
 & \multicolumn{3}{c|}{$\varepsilon$-based} & \multicolumn{3}{c|}{$\alpha$-based}\\
\hline 
Molecule  & 1$^{\rm st}$ best ($N_{T}$) & 2$^{\rm nd}$ best ($N_{T}$) & 3$^{\rm nd}$ best ($N_{T}$) & 1$^{\rm st}$ best ($N_{T}$) & 2$^{\rm nd}$ best ($N_{T}$) & 3$^{\rm nd}$ best ($N_{T}$)\\
\hline 
H$_{2}$  & QWC LF $(6.2\times 10^{6})$& QWC SI $(6.2\times 10^{6})$& FC LF $(6.2\times 10^{6})$& QWC SI $(1.4\times 10^{7})$& FC SI $(1.4\times 10^{7})$& QWC LF $(1.5\times 10^{7})$\\
\hline 
LiH  & QWC SI $(2.5\times 10^{8})$& FC SI $(2.7\times 10^{8})$& FC LF $(3.0\times 10^{8})$& FC SI $(2.9\times 10^{9})$& QWC SI $(2.9\times 10^{9})$& FC LF $(4.7\times 10^{9})$\\
\hline 
BeH$_{2}$  & FC SI $(5.5\times 10^{8})$& QWC SI $(6.2\times 10^{8})$& QWC LF $(7.0\times 10^{8})$& FC SI $(6.3\times 10^{9})$& QWC SI $(6.4\times 10^{9})$& QWC LF $(1.3\times 10^{10})$\\
\hline 
H$_{2}$O  & FC SI $(5.1\times 10^{8})$& QWC SI $(6.7\times 10^{8})$& QWC LF $(7.7\times 10^{8})$& FC SI $(5.9\times 10^{10})$& QWC SI $(6.0\times 10^{10})$& QWC LF $(1.0\times 10^{11})$\\
\hline 
NH$_{3}$  & QWC SI $(9.8\times 10^{8})$& QWC LF $(1.0\times 10^{9})$& FC SI $(1.2\times 10^{9})$& FC SI $(4.4\times 10^{10})$& QWC SI $(4.5\times 10^{10})$& QWC LF $(8.1\times 10^{10})$\\
\hline 
\end{tabular}
}

\caption{Best resource-efficient Hamiltonian decomposition methods 
for eigenvalue estimation within chemical accuracy with a Trotterized QPE algorithm. T-gate count $N_{T}$ is given 
in parenthesis.}
\label{Tgate_UBs}
\end{table*}

\begin{table*}
\centering
\begin{tabular}{|c|c|c|c|}
\hline 
 & \multicolumn{3}{c|}{$\varepsilon_{\text{app}}$-based}\\
\hline 
Molecule & 1$^{\rm st}$ best ($N_{T}$) & 2$^{\rm nd}$ best ($N_{T}$) & 3$^{\rm nd}$ best ($N_{T}$)\\
\hline 
H$_{2}$ & QWC LF $(6.2\times 10^{6})$& QWC SI $(6.2\times 10^{6})$& FC LF $(6.2\times 10^{6})$\\
\hline 
LiH & QWC SI $(2.5\times 10^{8})$& FC SI $(2.7\times 10^{8})$& FC LF $(3.0\times 10^{8})$\\
\hline 
BeH$_{2}$ & FC SI $(5.6\times 10^{8})$& QWC SI $(6.2\times 10^{8})$& QWC LF $(7.0\times 10^{8})$\\
\hline 
H$_{2}$O & FC SI $(7.6\times 10^{8})$& QWC SI $(8.8\times 10^{8})$& QWC LF $(9.7\times 10^{8})$\\
\hline 
NH$_{3}$ & FC SI $(1.0\times 10^{9})$& QWC LF $(1.1\times 10^{9})$& QWC SI $(1.2\times 10^{9})$\\
\hline 
\end{tabular}
\caption{Best resource-efficient Hamiltonian decomposition methods 
for eigenvalue estimation within chemical accuracy with a Trotterized QPE algorithm. T-gate count $N_{T}$ is given in 
parenthesis.}
\label{Ept_Table}
\end{table*}

Table \ref{Tgate_UBs} summarizes upper bound estimations of
the T-gate count required for QPE under a target accuracy based on
the exact scaling of Trotter approximation errors $\varepsilon$, alongside cruder
estimations based on the $\alpha$ upper bounds. Even though upper bound estimations on T-gate count based on the $\alpha$ mostly predict best performance of qubit decompositions, they tend to consistently overrate the FC SI and QWC SI methods. 
$\alpha$ based T gate count overestimates T gates by more than an order of magnitude for H$_2$O and NH$_3$. The overestimation is expected to grow as the sizes of molecules or the basis sets increase, since this leads to larger norm of the Hamiltonian and its fragments, and hence the error operators in Eq. \eqref{eq:alpha}. Thus, using $\alpha$'s leads to drastic overestimations of resources needed to obtain energies using the Trotter approximation and QPE. 

For $\varepsilon$ analysis based on perturbation theory expressions we cannot establish the same trends as those found for $\alpha$'s in Ref.~\cite{MartinezMartinez2023assessmentofvarious}.
Finally, in Table \ref{Ept_Table} we explore the faithfulness of $\varepsilon_{\text{app}}$ in the discrimination of the best resource-efficient methods, from which we note that $\varepsilon_{\text{app}}$-based estimator accurately captures the right order of magnitude of the T-gate numbers as obtained by the $\varepsilon$. Also, the $\varepsilon_{\text{app}}$-based estimator correctly suggests qubit partition methods as the most accurate compared to their fermionic counterparts. Due to similarity of T-gate numbers for various 
qubit partitionings, the ranking based on $\varepsilon_{\text{app}}$ and $\varepsilon$ are different, in spite of the high degree of $\varepsilon_{\text{app}}-\varepsilon$ correlation (Fig.\ref{fig_sim}). Since all the best Hamiltonian partitioning methods have very similar resource estimations and their particular order is of little importance, $\varepsilon_{\text{app}}$ can be a good substitute for $\varepsilon$.  \textcolor{black}{Thus, estimators of Trotter error based on perturbative expression [Eq. (\ref{eq:epsGS})] and a classically-accessible approximation to the electronic ground state, provide better correlation than commutator-norm-based counterparts, even for the strongly correlated molecular configurations.}

\section{Conclusions\textbf{\label{Conclusions}}}

We have calculated exact errors associated with the second order Trotter approximation for small molecules and different Hamiltonian partitionings. Previously derived commutator norm based upper bound, $\alpha$, was shown to have low correlation with the induced exact error in energy due to Trotter approximation. This confirmed the loose character of the $\alpha$ based upper bounds for energies, which makes these upper bounds inadequate in determining the true resources needed to achieve target accuracy in energy. The alternative estimate of the Trotter approximation error, $\varepsilon_{\text{app}}$, based on perturbative analysis of the effective Hamiltonian eigen-spectrum performed much better. The T gate upper bound estimates based on $\alpha$ were orders of magnitude higher than those predicted by the exact Trotter error. However, estimates based on $\varepsilon_{\text{app}}$ produced correct order of T gate estimates.\\

Substituting the exact ground eigenstate with a classically easy to obtain counterpart in calculating perturbation corrections gave accurate approximations, even in the case of strongly correlated molecular configurations. Specifically, the method produced accurate results in the case of H$_2$O and NH$_3$, where the CISD ground states have overlaps $93 \%$  and  $83 \%$ with respect to the exact ground state, respectively. For electronic systems with a higher degree of multiconfigurational character, one can find approximations to the global ground state using more sophisticated polynomial-in-time scaling methods, and hence make use of the tools developed here.  
These estimations of the Trotter approximation error raise two  
questions for future research: 1) how to 
optimize efficiently the Hamiltonian partitioning and ordering of its fragments 
based on the obtained error estimates; and 2) how to obtain upper bounds instead of approximations for 
the error estimates based on the eigen-spectrum analysis of $H_{\rm eff}$. 
Answering the second question will allow one to set an optimal Trotter time step for resource efficient simulation under a target energy eigenvalue estimation accuracy.

\section*{Data availability}
The code to generate the Hamiltonian fragments and calculate the exact and approximate Trotter errors can be found at https://github.com/Shashank-G-M/Perturbative\_Trotter\_Error. The same has been archived on Zenodo with DOI: https://zenodo.org/records/15327942.

\section*{Acknowledgments}

The authors would like to thank Nathan Wiebe for useful discussions.
L.A.M.M. is grateful to the Center for Quantum Information and Quantum
Control (CQIQC) for a postdoctoral fellowship.
P.D.K. is grateful to Mitacs for the Globalink research award. 
A.F.I. acknowledges financial support from the Natural Sciences 
and Engineering Council of Canada (NSERC). This research was partly enabled by the support of Compute Ontario (computeontario.ca) and the Digital Research Alliance of Canada (alliancecan.ca). Part of the computations were performed on the Niagara supercomputer at the SciNet HPC Consortium. SciNet is funded by Innovation, Science, and Economic Development Canada, the Digital Research Alliance of Canada, the Ontario Research Fund: Research Excellence, and the University of Toronto. 

\bibliographystyle{quantum}
\bibliography{citation}

\appendix

\section{Fermionic and Qubit-based Hamiltonian Decomposition methods\textbf{\label{Meths}}}

\label{Methods} Here, we discuss the methods we used to decompose
electronic Hamiltonians into fast-forwardable fragments using fermionic-
and qubit-based methods. The second quantized representation of the molecular
electronic Hamiltonian with $N$ single particle spin-orbitals under
this representation is 
\begin{equation}
\hat{H}=\sum_{pq=1}^{N}h_{pq}\hat{a}_{p}^{\dag}\hat{a}_{q}+\sum_{pqrs=1}^{N}g_{pqrs}\hat{a}_{p}^{\dag}\hat{a}_{q}\hat{a}_{r}^{\dag}\hat{a}_{s}\label{Hel}
\end{equation}
where $a_{p}^{\dag}$ $(a_{q})$ is the creation (annihilation) fermionic
operator for the $p^{th}$ spin-orbital, $h_{pq}$ and $g_{pqrs}$
are one- and two-electron integrals.\cite{olsen_bible}

\subsection{Fermionic partitioning methods}

These partitioning methods are built upon the solvability of one-electron
Hamiltonians using orbital rotations, according to 
\begin{equation}
\hat{H}_{1e}=\sum_{pq}h_{pq}\hat{a}_{p}^{\dag}\hat{a}_{q}=\hat{U}_{1}^{\dag}\left(\sum_{p}\tilde{h}_{p}\hat{n}_{p}\right)\hat{U}_{1},\label{solv_one}
\end{equation}
\begin{equation}
\hat{U}_{1}=\prod_{p>q}e^{\theta_{pq}(\hat{a}_{p}^{\dag}\hat{a}_{q}-\hat{a}_{q}^{\dag}\hat{a}_{p})}
\end{equation}
where $\hat{n}_{p}=\hat{a}_{p}^{\dag}\hat{a}_{q}$ occupation number
operators, $\tilde{h}_{p}$ are real constants, and $\hat{U}_{1}$
is an orbital rotation parameterized by the amplitudes $\theta_{pq}$.
Orbital rotations can also be employed to solve two-electron Hamiltonians
that are squares of one-electron Hamiltonians as follows: 
\begin{align}
\hat{H}^{(LR)} & =\left(\sum_{pq}h_{pq}\hat{a}_{p}^{\dagger}\hat{a}_{q}\right)^{2}=\hat{U}^{\dagger}\left(\sum_{p}\tilde{h}_{p}\hat{n}_{p}\right)^{2}\hat{U}\label{LRFrag}\\
 & =\hat{U}^{\dagger}\left(\sum_{pq}\tilde{h}_{p}\tilde{h}_{q}\hat{n}_{p}\hat{n}_{q}\right)\hat{U}.
\end{align}
The matrix with entries $\lambda_{pq}=\tilde{h}_{p}\tilde{h}_{q}$
is a rank-deficient one. The form of two-electron solvable Hamiltonians
by means of orbital rotations in (\ref{LRFrag}) can be straightforwardly
generalized by lifting the rank-deficient character of $\lambda$
matrix and regarding it as a full-rank hermitian matrix: 
\begin{equation}
\hat{H}^{(FR)}=\hat{U}^{\dagger}\left(\sum_{pq}\lambda_{pq}\hat{n}_{p}\hat{n}_{q}\right)\hat{U}.\label{FR_Frag}
\end{equation}
The fermionic methods that follow are classified according to whether
the Hamiltonian decomposition yields fast-forwardable fragments with
low- or full-rank character.

\textbf{Greedy Full Rank Optimization (GFRO)}: The approach uses orbital rotations to diagonalize the one-electron
part and approximate the two-body interaction terms featured in Eq.
(\ref{Hel}) as a sum of full-rank Hamiltonian fragments of the
form (\ref{FR_Frag}) \cite{PRXQuantum.2.040320} 
\begin{equation}
\hat{H}=\hat{H}_{1e}+\sum_{m=2}^{M}\hat{H}_{m}^{(FR)}.
\end{equation}
The decomposition is carried out in a greedy fashion to select an optimal Hamiltonian fragment $\hat{H}_{i+1}^{(FR)}$ [Eq. \eqref{FR_Frag}]
that minimizes the $L_{1}$ norm of the \textbf{$\tilde{G}$}$^{(i+1)}$
tensor at the $i^{th}$ iteration:
\begin{equation}
\sum_{pqrs=1}^{N}\tilde{G}_{pqrs}^{(i+1)}\hat{a}_{p}^{\dag}\hat{a}_{q}\hat{a}_{s}^{\dag}a_{s}=\sum_{pqrs}^{N}\tilde{G}_{pqrs}^{(i)}\hat{a}_{p}^{\dag}\hat{a}_{q}\hat{a}_{r}^{\dag}\hat{a}_{s}-\hat{H}_{i+1}^{(FR)}
\end{equation}
for $i\geq1$ and $\tilde{G}_{pqrs}^{(1)}=g_{pqrs}$, as a function of parameters $\{\lambda_{pq}^{(m)}\}$ and $\{\theta ^{(m)}\}$.


\textbf{Low-rank (LR) decomposition}: This partitioning method is
based on regarding the two-electron integral tensor $g_{pqrs}$ in
Eq. (\ref{Hel}) as a square matrix with composite indices along each
dimension. It has been shown \cite{Motta_2021} that rank-deficient
Hamiltonian fragments can be efficiently found by means of nested
factorizations on this matrix, such that 
\begin{equation}
\hat{H}=\hat{H}_{1e}+\sum_{m=2}^{M}\hat{H}_{m}^{(LR)},
\end{equation}
where 
\begin{equation}
\hat{H}_{m}^{(LR)}=\hat{U}_{m}^{\dag}\left(\sum_{p,q}\tilde{h}_{p}^{(m)}\tilde{h}_{q}^{(m)}\hat{n}_{p}\hat{n}_{q}\right)\hat{U}_{m}
\end{equation}
\\
\textbf{Pre- and post-processing of Hamiltonian fragments:} So far, the one-body electronic terms of the Hamiltonian in Eq. (\ref{Hel})
have been relegated given their straightforward orbital-rotation solvability.
However, the one-electron Hamiltonian in (\ref{Hel}) can be partitioned
in the same footing as the discussed methods by merging the former
in the two-body electronic terms as follows 
\begin{align}
\hat{H} & =\hat{U}_{1}^{\dagger}\left(\sum_{p}\varepsilon_{p}\hat{a}_{p}^{\dagger}\hat{a}_{p}\right)\hat{U}_{1}+\sum_{pqrs}g_{pqrs}\hat{a}_{p}^{\dagger}\hat{a}_{q}\hat{a}_{r}^{\dagger}\hat{a}_{s}\\
 & =\hat{U}_{1}^{\dag}\left(\sum_{pq,rs}[\tilde{g}_{pq,rs}+\varepsilon_{p}\delta_{pq}\delta_{pr}\delta_{ps}]\hat{a}_{p}^{\dag}\hat{a}_{q}\hat{a}_{r}^{\dag}\hat{a}_{s}\right)\hat{U}_{1}\\
 & =\sum_{p'q,r's'}\bar{g}_{p'q',r's'}\hat{a}_{p}^{\dag}\hat{a}_{q}\hat{a}_{r}^{\dag}\hat{a}_{s},\label{SD_Ham}
\end{align}
the decomposition of the ensuing two-electron Hamiltonian can be carried
out with the fermionic techniques discussed above. For computational
ease, in this work we consider the decomposition of the Hamiltonian
(\ref{SD_Ham}) with the GFRO approach, and refer to our combined
scheme as SD GFRO, where SD stands for "singles and doubles" in
analogy to the terminology used in the electronic structure literature
for single and double fermionic excitation operators. In addition
to the pre-processing discussed above, we consider a post-processing
technique that usually lowers the Trotter approximation error estimator $\alpha$
and relies on the removal of the one-body electron contributions encoded
within each of the two-body Hamiltonian fragments and grouping the
former in a single one-body electronic sub-Hamiltonian. This is accomplished
by employing the approach based in \cite{Lee_2021}, where two-body
interaction terms are written as a Linear Combination of Unitaries
(LCU), with a concomitant adjustment of the one-body Hamiltonian contributions:
\cite{MartinezMartinez2023assessmentofvarious} 
\begin{align}
\hat{H} & =\sum_{pq=1}^{N}\left(h_{pq}+\sum_{l}g_{pq,ll}\right)\hat{a}_{p}^{\dag}\hat{a}_{q}+\sum_{l=2}\hat{U}_{l}^{\dag}\left(\sum_{i,j}^{N}\frac{\lambda_{ij}^{(l)}}{4}\hat{r}_{i}\hat{r}_{i}\right)\hat{U}_{l}\\
 & -\frac{1}{4}\sum_{p,q}g_{pp,qq}.
\end{align}

\subsection{Qubit-based partitioning methods}

When the Hamiltonian (\ref{Hel}) is mapped to $N$ interacting two-level
systems through encodings such as Jordan-Wigner or Bravyi-Kitaev,
the Hamiltonian thus obtained is of the form, 
\begin{equation*}
\hat{H}_{q}=\sum_{n}{c}_{n}\hat{P}_{n}\qquad\text{where}\qquad\hat{P}_{n}=\otimes_{k=1}^{N}\hat{\sigma}_{k}^{(n)}\label{2-FH_hamiltonian}
\end{equation*}
where, $c_{n}$ are numerical coefficients and $\hat{P}_{n}$ are
tensor products of single-qubit Pauli operators and the identity,
$\hat{\sigma}_{k}^{(n)}=\hat{x}_{k},\hat{y}_{k},\hat{z}_{k},\hat{I}_{k}$,
acting on the $k^{th}$ qubit. The Fully Commuting (FC) grouping partitions
$\hat{H}_{q}$ into $\hat{H}_{n}^{(FC)}$ fragments containing commuting
Pauli products: 
\begin{equation*}
\text{if}\qquad\hat{P}_{i},\hat{P}_{j}\in\hat{H}_{n}^{(FC)}\qquad\text{then}\qquad[\hat{P}_{i},\hat{P}_{j}]=0.
\end{equation*}
This FC condition ensures that $\hat{H}_{n}^{(FC)}$ can be transformed,
through a series of Clifford group transformations, into sums of only
products of Pauli $\hat{z}_{k}$ operators.\cite{bansingh2022fidelity,van2020}
We also consider a grouping with a more strict condition known as
qubit-wise commutativity (QWC), where each single-qubit Pauli operator
in one product commutes with its counterpart in the other product.
For example, $\hat{x}_{1}\hat{y}_{2}\hat{I}_{3}$ and $\hat{x}_{1}\hat{I}_{2}\hat{z}_{3}$
have QWC as $[\hat{x}_{1},\hat{x}_{1}]=0$, $[\hat{y}_{2},\hat{I}_{2}]=0$,
$[\hat{I}_{3},\hat{z}_{3}]=0$. Hence, both terms must also fully
commute. The converse does not always hold true. For example, $\hat{x}_{1}\hat{x}_{2}$
and $\hat{y}_{1}\hat{y}_{2}$ are fully commuting but not qubit-wise
commuting.\cite{yen2020measuring}

For the FC and QWC partitioning techniques, we work with the largest-first
(LF) heuristic and the Sorted Insertion (SI) algorithm. The SI algorithm
is based on a greedy partitioning of the Hamiltonian, which results
in concentrated coefficients $c_{n}$ in the first found Hamiltonian
fragments. The LF algorithm, in contrast, yields a homogeneous distribution
in the magnitudes of the $c_{n}$ coefficients across Hamiltonian
fragments, which usually results in a smaller number of fragments
compared to the SI version \cite{Verteletskyi_2020,yen2020measuring}.


\subsection{Details of the Hamiltonians and Wavefunctions}

The Hamiltonians were generated using the STO-3G basis and the Jordan-Wigner
transformations for qubit encodings as implemented in the OpenFermion
package \cite{mcclean2020openfermion}. The nuclear geometries for the molecules are given by R(H-H)=1
Å (H$_{2}$), R(Li-H)=1 Å (LiH) and R(Be-H)=1 Å with collinear atomic
arrangement (BeH$_{2}$), R(OH)= 1.9 Å and $\angle HOH$=104.5$^\circ$
(H$_{2}$O); and R(N-H)=1.9 Å with $\angle HNH$=107$^\circ$ (NH$_{3}$).
The ground state CISD wavefunction is generated using the OpenFermion package.









\subsection{Computation of errors for the second order Trotter approximation
\label{AutoCorr}}

From Eq. (\ref{eq:Heff}) of the main text, $\hat{H}_{\text{eff}}$ is computed through 
\begin{equation}
    \hat{H}_{\text{eff}}=it^{-1}\ln\left(\hat{U}_{T}^{(2)}(t)\right),
\end{equation}
where $t=\mathcal{O}(|| \hat{H} ||^{-1})$. $\varepsilon$'s are obtained according to $\varepsilon=t^{-2}|E^{(T)}_{0}-E_{0}|$, where $E^{(T)}_{0}$ ($E_{0}$) is the ground state energy of $\hat{H}_{\text{eff}}$ ($\hat{H}$). All these calculations were performed using the python Scipy library \cite{SciPy}.
To reduce computational overhead in our calculations, we take advantage
of the fact that the initial state $|\psi\rangle$ belongs to a particular 
irreducible representation of the molecular symmetries: the number of electrons, 
$\hat N_e$, the electron spin, $\hat{S}^{2}$, and its projection, $\hat{S}_{z}$.
Selecting symmetry adapted states for the neutral singlet molecular forms allowed 
to reduce the Hamiltonian sub-spaces by almost two orders of magnitude. Similarly, for qubit-based partitioning methods, we use qubit tapering
to reduce the system size of NH$_{3}$ from a 16-qubit system to a
14-qubit system \cite{bravyi2017tapering}. Since the number qubit fragments and their sizes are usually large for BeH$_2$, H$_2$O and NH$_3$, instead of exponentiating each fragment exactly, we approximate the exponential using Taylor series up to 11$^{\text{th}}$ order in time. We make use of the Niagara compute cluster hosted by SciNet \cite{niagara} for memory intensive calculations. The Trotter approximation error depends on the order in which individual
unitaries $e^{-it\hat{H}_{n}}$ are applied \cite{Tranter2019}. The code to generate Hamiltonian fragments and calculate the Trotter error can be accessed at https://doi.org/10.5281/zenodo.15327942.

\subsection{Effective Hamiltonian  derivation based on BCH expansion.\label{Eff_BCH}}


In this section we generalize the BCH formula, usually defined for two Hamiltonian fragments, to an arbitrary number of fragments $N$. We will use mathematical induction with a starting point:
\begin{equation}
e^{-iH_{2}t}e^{-iH_{1}t}=\exp\left(-iH_{\rm eff}^{(2)}t\right),\label{Heff_2}
\end{equation}
where 
\[
H_{\rm eff}^{(2)}=H_{2}+H_{1}+\frac{(-i)}{2}t[H_{2},H_{1}]+\frac{(-i)^{2}}{12}t^{2}[H_{2},[H_{2},H_{1}]]-\frac{(-i)^{2}}{12}t^{2}[H_{1},[H_{2},H_{1}]]+\mathcal{O}(t^{3}).
\]
To obtain the form of the effective Hamiltonian for $N$ fragments, $H^{(N)}_{\text{eff}}$ we extend Eq. (\ref{Heff_2}) to the three-fragment case:
\begin{align*}
e^{-iH_{3}t}e^{-iH_{2}t}e^{-iH_{1}t} & =e^{-iH_{3}t}e^{-iH_{\rm eff}^{(2)}t}=\exp\bigg(-iH_{3}t-iH_{\rm eff}^{(2,1)}t+\frac{(-i)^{2}}{2}t^{2}[H_{3},H_{\rm eff}^{(2,1)}]\\
 & +\frac{(-i)^{3}}{12}t^{3}[H_{3},[H_{3},H_{\rm eff}^{(2,1)}]]-\frac{(-i)^{3}}{12}t^{3}[H_{\rm eff}^{(2,1)},[H_{3},H_{\rm eff}^{(2,1)}]]+\mathcal{O}(t^{4})\bigg)\\
 & =\exp\left(\hat{A}\right)
\end{align*}
where $\hat{A}$ becomes
\begin{align*}
\hat{A} & =-iH_{3}t-iH_{2}t-iH_{1}t+\frac{(-i)^{2}}{2}t^{2}[H_{2},H_{1}]+\frac{(-i)^{2}}{2}t^{2}[H_{3},H_{1}]+\frac{(-i)^{2}}{2}t^{2}[H_{3},H_{2}]\\
 & +\frac{(-i)^{3}}{12}t^{3}[H_{2},[H_{2},H_{1}]]-\frac{(-i)^{3}}{12}t^{3}[H_{1},[H_{2},H_{1}]]+\frac{(-i)^{3}}{4}t^{3}[H_{3},[H_{2},H_{1}]]\\
 & +\frac{(-i)^{3}}{12}t^{3}[H_{3},[H_{3},H_{2}]]+\frac{(-i)^{3}}{12}t^{3}[H_{3},[H_{3},H_{1}]]-\frac{(-i)^{3}}{12}t^{3}[H_{1},[H_{3},H_{1}]]\\
 & -\frac{(-i)^{3}}{12}t^{3}[H_{2},[H_{3},H_{1}]]-\frac{(-i)^{3}}{12}t^{3}[H_{1},[H_{3},H_{2}]]-\frac{(-i)^{3}}{12}t^{3}[H_{2},[H_{3},H_{2}]]+\mathcal{O}(t^{4})\\
 & =-iH_{3}t-iH_{2}t-iH_{1}t+\frac{(-i)^{2}}{2}t^{2}\sum_{v>\mu}^{3}[H_{v},H_{\mu}]+\frac{(-i)^{3}}{4}t^{3}\sum_{v'>v>\mu}^{3}[H_{v'},[H_{v},H_{\mu}]]\\
 & +\frac{(-i)^{3}}{12}t^{3}\sum_{v>\mu}^{3}[H_{v},[H_{v},H_{\mu}]]-\frac{(-i)^{3}}{12}t^{3}\sum_{v>\mu,v'}^{3}[H_{v'},[H_{v},H_{\mu}]]+\mathcal{O}(t^{4}).
\end{align*}
We note that $\hat{A}$ can be written in the form
\begin{align}\label{eq:Gen_A}
\hat{A} & =-it\left(H^{(3)}+\frac{t}{2}\hat{v}_{1}^{(3)}+\frac{t^{2}}{3}\hat{v}_{2}^{(3)}+i\frac{t^{2}}{12}[H^{(3)},\hat{v}_{1}^{(3)}]+\mathcal{O}(t^{3})\right),
\end{align}
where
\begin{align*}
H^{(n)} & =\sum_{j=1}^{n}H_{j},\\
\hat{v}_{1}^{(n)} & =-i\sum_{v=\mu+1}^{n}\sum_{\mu=1}^{n-1}[H_{v},H_{\mu}],\\
\hat{v}_{2}^{(n)} & =-\sum_{v'=v}^{n}\sum_{v=\mu+1}^{n}\sum_{\mu=1}^{n-1}\left(1-\frac{\delta_{v',v}}{2}\right)[H_{v'},[H_{v},H_{\mu}]].
\end{align*}
Finally, to show that the form (\ref{eq:Gen_A}) can be generalized for an arbitrary number of Hamiltonian fragments, we use induction:
\begin{align*}
e^{-iH_{n+1}t}e^{-iH_{\rm eff}^{(n)}t} & =\exp\bigg(-iH_{\rm eff}^{(n)}t-iH_{n+1}t+\frac{(-i)^{2}}{2}t^{2}[H_{n+1},H_{\rm eff}^{(n)}]+\frac{(-i)^{3}}{12}t^{3}[H_{n+1},[H_{n+1},H_{\rm eff}^{(n)}]]\\
 & -\frac{(-i)^{3}}{12}t^{3}[H_{\rm eff}^{(n)},[H_{n+1},H_{\rm eff}^{(n)}]]+\mathcal{O}(t^{4})\bigg)\\
 & =\exp\left(\hat{B}\right),
\end{align*}
where
\begin{align*}
\hat{B} & =-iH_{n+1}t-iH^{(n)}t-i\frac{t^{2}}{2}\hat{v}_{1}^{(n)}-i\frac{t^{3}}{3}\hat{v}_{2}^{(n)}+\frac{t^{3}}{12}[\hat{H}^{(n)},\hat{v}_{1}^{(n)}]\\
 & +\frac{(-i)^{2}}{2}t^{2}[H_{n+1},H^{(n)}+\frac{t}{2}\hat{v}_{1}^{(n)}]+\frac{(-i)^{3}}{12}t^{3}[H_{n+1},[H_{n+1},H^{(n)}]]\\
 & -\frac{(-i)^{3}}{12}t^{3}[H^{(n)},[H_{n+1},H^{(n)}]]+\mathcal{O}(t^{4})\\
 & =-i\left(H_{n+1}+H^{(n)}\right)t-i\frac{t^{2}}{2}\left(\hat{v}_{1}^{(n)}-i[H_{n+1},H^{(n)}]\right)\\
 & -i\frac{t^{3}}{3}\left(\hat{v}_{2}^{(n)}-\frac{1}{2}[H_{n+1},[H_{n+1},H^{(n)}]]-i[H_{n+1},\hat{v}_{1}^{(n)}]\right)\\
 & +\frac{t^{3}}{12}\bigg([H^{(n)},\hat{v}_{1}^{(n)}]-i[H^{(n)},[H_{n+1},H^{(n)}]]+[H_{n+1},\hat{v}_{1}^{(n)}]\\
 & -i[H_{n+1},[H_{n+1},H^{(n)}]]\bigg).
\end{align*}
By using
\[
H^{(n+1)}=H^{(n)}+H_{n+1}
\]
\[
\hat{v}_{1}^{(n+1)}=\hat{v}_{1}^{(n)}-i[\hat{H}_{n+1},\hat{H}^{(n)}]
\]
\[
\hat{v}_{2}^{(n+1)}=\hat{v}_{2}^{(n)}-\frac{1}{2}[H_{n+1},[H_{n+1},H^{(n)}]]-i[H_{n+1},\hat{v}_{1}^{(n)}]
\]
\begin{align*}
[\hat{H}^{(n+1)},\hat{v}_{1}^{(n+1)}] & =[H^{(n)},\hat{v}_{1}^{(n)}]-i[H^{(n)},[H_{n+1},H^{(n)}]]\\
 & +[H_{n+1},\hat{v}_{1}^{(n)}]-i[H_{n+1},[H_{n+1},H^{(n)}]]
\end{align*}
we have
\[
H_{\rm eff}^{(n+1)}=H^{(n+1)}+\frac{t}{2}\hat{v}_{1}^{(n+1)}+\frac{t^{2}}{3}\hat{v}_{2}^{(n+1)}+i\frac{t^{2}}{12}[H^{(n+1)},\hat{v}_{1}^{(n+1)}]+\mathcal{O}(t^{3}).
\]
Therefore, for Hamiltonian $H$ decomposed into $N$ Hamiltonian fragments, the effective Hamiltonian $H_{\text{eff}}$ is
\begin{equation}
\begin{split}
    H_{\text{eff}}&=H^{(N)}+\frac{\tau}{2}\hat{v}^{(N)}_{1}+\frac{\tau^{2}}{3}\hat{v}^{(N)}_{2}+i\frac{\tau^{2}}{12} [H^{(N)},\hat{v}^{(N)}_{1}]+\mathcal{O}(\tau^{3})\\
    &=H+\frac{\tau}{2}\hat{v}_{1}+\frac{\tau^{2}}{3}\hat{v}_{2}+i\frac{\tau^{2}}{12} [H,\hat{v}_{1}]+\mathcal{O}(\tau^{3})\\
    &=H+\hat{V}_{1}\tau+\hat{V}_{2}\tau^{2}+\mathcal{O}(\tau^{3}),
    \end{split}
\end{equation}
where
\begin{equation}
    \begin{split}
        \hat{V}_{1}&=\frac{\hat{v}_{1}}{2}=-\frac{i}{2}\sum_{v=\mu+1}^{N}\sum_{\mu=1}^{N-1}[H_{v},H_{\mu}],\\
        \hat{V}_{2} &= \frac{1}{3}\hat{v}_{2}+\frac{i}{12}[H,\hat{v}_{1}]\\
        &= -\frac{1}{3}\sum_{v'=v}^{N}\sum_{v=\mu+1}^{N}\sum_{\mu=1}^{N-1}\left(1-\frac{\delta_{v',v}}{2}\right)[H_{v'},[H_{v},H_{\mu}]]+\frac{i}{6}[H,\hat{V}_{1}].
    \end{split}
\end{equation}

\textcolor{black}{To get the special case of second order Trotter, use $N = 2M$, where $M$ is the number of Hamiltonian fragments, and $H_{M+i}$ = $H_{M+1-i}$ for $i=1$ to $M$. Also, each of the fragment will have to be rescaled by a factor of half, as we repeat each fragment twice in the second order Tortter formula [see Eq. \eqref{trotter_2nd}]. With this constraint, for every commutator $[H_{\mu}, H_{\nu}]$ in the expression of $\hat{V}_1$, there exists a commutator $[H_{\nu}, H_{\mu}]$ with the same coefficient. Thus, $\hat{V}_1$ equals zero. Using the same constraint in the expression of $\hat{V}_2$, we recover Eq. \eqref{Pert_defs}}.

\subsection{Compendium of different Trotter approximation error upper bounds}\label{EpsTables}

Tables \ref{II-1}-\ref{VI-1} compile Trotter approximation error estimates 
based on $\varepsilon$, $\alpha$, and $\alpha_e$ quantities. 
Tables \ref{C2_HF} and \ref{C2_exact} summarize $\varepsilon_{\text{app}}$ and $\varepsilon_{2}$ values. These results are obtained by considering the Trotterized unitary:
\begin{align*}
\hat{U}^{(2)}_{T}(t) = \prod_{m=1}^{M}e^{-i\hat{H}_{m}t/2} \prod_{m=M}^{1}e^{-i\hat{H}_{m}t/2}.
\end{align*}
where the ordering of Hamiltonian fragments was taken as found by the different partition methods with no further post-processing.


\begin{table*}
\centering %
\resizebox{\textwidth}{!}{
\begin{tabular}{||c|c|c|c|c|c|c|c|c|c|}
\hline 
Molecule & QWC & QWC & FC & FC & LR & GFRO & LR & GFRO & SD\\
 & LF & SI & LF & SI & LCU & LCU &  &  & GFRO\\
\hline 
\hline 
H$_{2}$ & $3.3\times10^{-3}$ & $3.3\times10^{-3}$ & $3.3\times10^{-3}$ & $3.3\times10^{-3}$ & $3.3\times10^{-3}$ & $3.3\times10^{-3}$ & $2.8\times10^{-3}$ & $2.8\times10^{-3}$ & $3.1\times10^{-3}$ \\
\hline
LiH & $3.2\times10^{-3}$ & $2.2\times10^{-3}$ & $3.0\times10^{-3}$ & $2.4\times10^{-3}$ & $3.3\times10^{-3}$ & $3.4\times10^{-3}$ & $4.7\times10^{-2}$ & $5.0\times10^{-2}$ & $1.8\times10^{-2}$ \\
\hline
BeH$_{2}$ & $1.4\times10^{-2}$ & $1.1\times10^{-2}$ & $2.3\times10^{-2}$ & $8.8\times10^{-3}$ & $9.3\times10^{-3}$ & $9.6\times10^{-3}$ & $2.9\times10^{-2}$ & $3.3\times10^{-2}$ & $2.0\times10^{-2}$ \\
\hline
H$_{2}$O & $6.2\times10^{-3}$ & $4.8\times10^{-3}$ & $2.4\times10^{-2}$ & $2.9\times10^{-3}$ & $2.4\times10^{-2}$ & $2.5\times10^{-2}$ & $1.4\times10^{-1}$ & $1.3\times10^{-1}$ & $2.6\times10^{-2}$ \\
\hline
NH$_{3}$ & $1.1\times10^{-2}$ & $1.0\times10^{-2}$ & $9.0\times10^{-2}$ & $1.5\times10^{-2}$ & $2.0\times10^{-2}$ & $2.0\times10^{-2}$ & $1.7\times10^{-1}$ & $1.4\times10^{-1}$ & $2.9\times10^{-2}$ \\
\hline
\end{tabular}
}
\caption{$\varepsilon$ values obtained from true Trotter approximation error scaling for different
fermionic and qubit-based partitioning methods and molecules}
\label{II-1}
\end{table*}

\begin{table*}
\centering %
\begin{tabular}{||c|c|c|c|c|c|c|c|c|c|}
\hline 
Molecule & QWC & QWC & FC & FC & LR & GFRO & LR & GFRO & SD \\
 & LF & SI & LF & SI & LCU & LCU &  &  & GFRO \\
\hline 
\hline 
H$_{2}$ & 0.02 & 0.02 & 0.02 & 0.02 & 0.02 & 0.02 & 0.02 & 0.02 & 0.02 \\
\hline
LiH & 1.07 & 0.26 & 0.63 & 0.26 & 0.13 & 0.12 & 0.52 & 0.46 & 0.23 \\
\hline
BeH$_{2}$ & 4.22 & 1.02 & 4.96 & 0.99 & 0.58 & 0.55 & 2.36 & 2.03 & 1.13 \\
\hline
H$_{2}$O & 79.41 & 28.73 & 181.56 & 27.86 & 15.30 & 15.06 & 52.37 & 48.27 & 27.88 \\
\hline
NH$_{3}$ & 51.66 & 16.36 & 65.99 & 16.02 & 7.81 & 7.64 & 28.43 & 25.79 & 14.51 \\
\hline
\end{tabular}\caption{Values of Trotter approximation error upper bound $\alpha$ as defined in Eq. \eqref{eq:alpha}.}
\label{V-1}
\end{table*}

\begin{table*}
\centering %
\begin{tabular}{||c|c|c|c|c|c|c|c|c|c|}
\hline 
Molecule & QWC & QWC & FC & FC & LR & GFRO & LR & GFRO & SD\\
 & LF & SI & LF & SI & LCU & LCU &  &  & GFRO\\
\hline 
\hline 
H$_{2}$ & 0.02 & 0.01 & 0.02 & 0.01 & 0.01 & 0.01 & 0.01 & 0.01 & 0.01 \\
\hline
LiH & 0.22 & 0.18 & 0.25 & 0.18 & 0.10 & 0.10 & 0.07 & 0.08 & 0.06 \\
\hline
BeH$_{2}$ & 0.67 & 0.75 & 0.81 & 0.76 & 0.42 & 0.43 & 0.29 & 0.35 & 0.36 \\
\hline
H$_{2}$O & 23.25 & 23.42 & 46.45 & 23.45 & 1.85 & 11.86 & 15.67 & 14.88 & 12.87 \\
\hline
 NH$_3$ & 11.23 & 11.23 & 15.86 & 11.21 & 6.55 & 6.56 & 7.25 & 7.05 & 6.03 \\
\hline
\end{tabular}\caption{Values of $\alpha_e=\norm{\hat{U}_T(t) - \hat{U}_T^{(2)}(t)}/t^{3}$.}
\label{VI-1}
\end{table*}

\begin{table*}[t]
\centering

\resizebox{0.9\textwidth}{!}{

\begin{tabular}{|c|c|c|c|c|c|c|l|l|l|}
\hline 
Molecule & QWC-LF  & QWC-SI  & FC-LF  & FC-SI  & LR LCU  & GFRO LCU & LR  & GFRO  &SD-GFRO 
\\
\hline 
H$_{2}$ & $3.24\times10^{-3}$ & $3.24\times10^{-3}$ & $3.24\times10^{-3}$ & $3.24\times10^{-3}$ & $3.24\times10^{-3}$ & $3.24\times10^{-3}$ & $2.77\times10^{-3}$ & $2.78\times10^{-3}$ & $3.00\times10^{-3}$ \\
\hline
LiH & $3.26\times10^{-3}$ & $2.18\times10^{-3}$ & $3.02\times10^{-3}$ & $2.46\times10^{-3}$ & $3.30\times10^{-3}$ & $3.39\times10^{-3}$ & $4.72\times10^{-2}$ & $4.99\times10^{-2}$ & $1.82\times10^{-2}$ \\
\hline
BeH$_{2}$ & $1.38\times10^{-2}$ & $1.12\times10^{-2}$ & $2.25\times10^{-2}$ & $8.93\times10^{-3}$ & $9.49\times10^{-3}$ & $9.83\times10^{-3}$ & $2.89\times10^{-2}$ & $3.36\times10^{-2}$ & $1.98\times10^{-2}$ \\
\hline
H$_{2}$O & $9.78\times10^{-3}$ & $8.04\times10^{-3}$ & $1.99\times10^{-2}$ & $6.03\times10^{-3}$ & $3.22\times10^{-2}$ & $3.52\times10^{-2}$ & $1.78\times10^{-1}$ & $1.67\times10^{-1}$ & $3.48\times10^{-2}$ \\
\hline
NH$_{3}$ & $1.31\times10^{-2}$ & $1.57\times10^{-2}$ & $7.79\times10^{-2}$ & $1.05\times10^{-2}$ & $3.33\times10^{-2}$ & $3.44\times10^{-2}$ & $2.34\times10^{-1}$ & $2.09\times10^{-1}$ & $4.75\times10^{-2}$ \\
\hline
\end{tabular}

}
\caption{$\varepsilon_{\text{app}} = \langle\psi_{0}|\hat{V}_{2}|\psi_{0}\rangle$ for different
Hamiltonian decomposition methods and molecules.}

\label{C2_HF}
\end{table*}

\begin{table*}
\centering

\resizebox{0.9\textwidth}{!}{

\begin{tabular}{|c|c|c|c|c|c|c|l|l|l|}
\hline 
Molecule & QWC-LF & QWC-SI & FC-LF & FC-SI & LR LCU & GFRO LCU & LR & GFRO &SD-GFRO 
\\
\hline 
H$_{2}$ & $3.24\times10^{-3}$ & $3.24\times10^{-3}$ & $3.24\times10^{-3}$ & $3.24\times10^{-3}$ & $3.24\times10^{-3}$ & $3.24\times10^{-3}$ & $2.77\times10^{-3}$ & $2.78\times10^{-3}$ & $3.00\times10^{-3}$ \\
\hline
LiH & $3.25\times10^{-3}$ & $2.16\times10^{-3}$ & $3.01\times10^{-3}$ & $2.45\times10^{-3}$ & $3.30\times10^{-3}$ & $3.39\times10^{-3}$ & $4.72\times10^{-2}$ & $4.99\times10^{-2}$ & $1.82\times10^{-2}$ \\
\hline
BeH$_{2}$ & $1.37\times10^{-2}$ & $1.10\times10^{-2}$ & $2.27\times10^{-2}$ & $8.76\times10^{-3}$ & $9.30\times10^{-3}$ & $9.63\times10^{-3}$ & $2.87\times10^{-2}$ & $3.34\times10^{-2}$ & $1.96\times10^{-2}$ \\
\hline
H$_{2}$O & $6.22\times10^{-3}$ & $4.76\times10^{-3}$ & $2.36\times10^{-2}$ & $2.85\times10^{-3}$ & $2.37\times10^{-2}$ & $2.51\times10^{-2}$ & $1.42\times10^{-1}$ & $1.28\times10^{-1}$ & $2.58\times10^{-2}$ \\
\hline
NH$_{3}$ & $1.15\times10^{-2}$ & $9.98\times10^{-3}$ & $8.95\times10^{-2}$ & $1.52\times10^{-2}$ & $2.00\times10^{-2}$ & $1.98\times10^{-2}$ & $1.65\times10^{-1}$ & $1.38\times10^{-1}$ & $2.94\times10^{-2}$ \\
\hline
\end{tabular}
}

\caption{$\varepsilon_2 = \langle\phi_{0}|\hat{V}_{2}|\phi_{0}\rangle$ for different
Hamiltonian decomposition methods and molecules.}

\label{C2_exact}
\end{table*}


\subsection{T-gate count upper bound estimations}
Upper-bound for T-gate counts for a fixed target
error $\varepsilon_{Tot}$ in energy eigenvalue estimation in a Trotterized
Quantum Phase Estimation algorithm can be formulated in light of previous
works \cite{Berry2009HowTP,Kivlichan2020improvedfault}. 
The total T-gate
count $N_{T}$ \cite{Reiher_2017,Kivlichan2020improvedfault} is given
by 
\begin{equation}
N_{T}=N_{R}N_{HT}N_{PE}
\end{equation}
where $N_{R}$ is the number of single-qubit rotations needed for
the implementation of a single Trotter step in a quantum computer.
$N_{HT}$ refers to the number of T gates needed to compile one single
qubit rotation (for a fixed target error $\varepsilon_{HT}$) and $N_{PE}$
is the number of Trotter steps required to resolve the target energy
eigenvalue under a target uncertainty $\varepsilon_{PE}$, the latter
scaling as $t^{-1}$, $t$ being the total simulation time. Using
our results that describe the energy deviation in the estimated ground-state
energy eigenvalue due to the Trotter approximation, according to the
relation $\varepsilon \Delta t^{2}=\Delta E_{T}$, we find the Trotter step $\Delta t$
according to a target error $\varepsilon_{TS}$, given by $\Delta t=\sqrt{\frac{\varepsilon_{TS}}{\varepsilon}}$.
The number of Trotter steps needed for a target uncertainty in phase
estimation under adaptive phase estimation techniques is given by
\begin{equation}
N_{PE}\approx\frac{0.76\pi}{\varepsilon_{PE}\Delta t}=\frac{0.76\pi\sqrt{\varepsilon}}{\varepsilon_{PE}\sqrt{\varepsilon_{TS}}}
\end{equation}
Finally, the number of T gates needed to compile one single qubit
rotation for a fixed target error $\varepsilon_{HT}$ is $N_{HT}=1.15\log_{2}\left(\frac{N_{R}}{\varepsilon_{HT}\Delta t}\right)+9.2=1.15\log_{2}\left(\frac{N_{R}\sqrt{\varepsilon}}{\varepsilon_{HT}\sqrt{\varepsilon_{TS}}}\right)$
+ 9.2. Putting everything together we arrive at 
\begin{equation}
N_{T}\approx\frac{0.76\pi N_{R}\sqrt{\varepsilon}}{\sqrt{\varepsilon_{TS}}\varepsilon_{PE}}\left[1.15\log_{2}\left(\frac{N_{R}\sqrt{\varepsilon}}{\varepsilon_{HT}\sqrt{\varepsilon_{TS}}|}\right)+9.2\right]\label{TightNT}
\end{equation}
In the worst case, the errors due to the three sources discussed above,
add linearly \cite{Kivlichan2020improvedfault} and to guarantee that
the total error is at most $\varepsilon_{Tot}$ we assume 
\begin{equation}
\varepsilon_{Tot}=\varepsilon_{TS}+\varepsilon_{PE}+\varepsilon_{HT}.\label{sum_errs}
\end{equation}
Thus, we can minimize the number of T-gates $N_{T}$ over the target
errors in Eq. (\ref{TightNT}) subject to the constraint (\ref{sum_errs}),
for an estimation of T-gate under a target error $\varepsilon_{Tot}$. In this work, we have taken $\varepsilon_{Tot} = 1.6 \times 10^{-3}$ Hartree, the chemical accuracy.

\clearpage 

\end{document}